\documentclass[12pt,preprint]{aastex}
\usepackage{lmcommands}

\shorttitle{\ion{H}{1} in Compact Groups}
\shortauthors{Walker et al.}

\begin{document}

\title{Global Properties of Neutral Hydrogen in Compact Groups}

\author{Lisa May Walker\altaffilmark{1}}
\affil{Steward Observatory, University of Arizona, Tucson, AZ 85721}

\author{Kelsey E. Johnson}
\affil{Department of Astronomy, University of Virginia, Charlottesville, VA 22904}

\author{Sarah C. Gallagher\altaffilmark{2}}
\affil{Department of Physics and Astronomy, University of Western Ontario, London, ON}

\author{George C. Privon\altaffilmark{1}}
\affil{Departamento de Astronom{\' i}a, Universidad de Concepci{\' o}n, Concepci{\' o}n, Chile}

\author{Amanda A. Kepley\altaffilmark{1}}
\affil{National Radio Astronomy Observatory, Charlottesville, VA 22903}

\author{David G. Whelan\altaffilmark{1}}
\affil{Physics Department, Austin College, Sherman, TX 75090}

\author{Tyler D. Desjardins\altaffilmark{3}}
\affil{Department of Physics and Astronomy, University of Kansas, Lawrence, KS 66045}

\author{Ann I. Zabludoff}
\affil{Department of Astronomy and Steward Observatory, University of Arizona, Tucson, AZ 85721}

\altaffiltext{1}{Department of Astronomy, University of Virginia, Charlottesville, VA 22904}
\altaffiltext{2}{Visiting Fellow, Yale Center for Astronomy and Astrophysics, Department of Physics, Yale University, New Haven, CT}
\altaffiltext{3}{Department of Physics and Astronomy, University of Western Ontario, London, ON}

\begin{abstract}
Compact groups of galaxies provide a unique environment to study the evolution of galaxies amid frequent gravitational encounters. These nearby groups have conditions similar to those in the earlier universe when galaxies were assembled and give us the opportunity to witness hierarchical formation in progress. To understand how the compact group environment affects galaxy evolution, we examine the gas and dust in these groups. We present new single-dish GBT neutral hydrogen (\ion{H}{1}) observations of 30 compact groups and define a new way to quantify the group \ion{H}{1} content as the \ion{H}{1}-to-stellar mass ratio of the group as a whole. We compare the \ion{H}{1} content with \mir indicators of star formation and optical $\left[g-r\right]$ color to search for correlations between group gas content and star formation activity of individual group members. Quiescent galaxies tend to live in \ion{H}{1}-poor groups, and galaxies with active star formation are more commonly found in \ion{H}{1}-rich groups. Intriguingly, we also find ``rogue'' galaxies whose star formation does not correlate with group \ion{H}{1} content. In particular, we identify three galaxies (NGC 2968 in RSCG 34, KUG 1131+202A in RSCG 42, and NGC 4613 in RSCG 64) whose \mir activity is discrepant with the \ion{H}{1}. We speculate that this mismatch between \mir activity and \ion{H}{1} content is a consequence of strong interactions in this environment that can strip \ion{H}{1} from galaxies and abruptly affect star-formation. Ultimately, characterizing how and on what timescales the gas is processed in compact groups will help us understand the interstellar medium in complex, dense environments similar to the earlier Universe.
\end{abstract}

\keywords{galaxies: groups: general --- galaxies: evolution --- galaxies: interactions --- galaxies: intergalactic medium --- galaxies: ISM --- ISM: general}

\section{INTRODUCTION}\label{intro}
Compact groups of galaxies, at the dense end of the group distribution, provide an opportunity to observe galaxies in an environment similar to that of the earlier universe \citep{baron87}. Additionally, the majority of galaxies in the local universe are found in a group environment \citep[e.g.,][]{tully87, small99, karachentsev05}, making groups essential to study in order to understand galaxy evolution. Galaxy interactions in groups are more complicated to characterize than in isolated merger systems, and evolutionary scenarios must take into account both the interstellar and intragroup media. Unlike clusters, properties of compact groups can evolve on rapid time scales \citep[$\sim400$~Myr;][]{borthakur10} and are often unvirialized \citep[e.g.,][]{desjardins13}.

Previous work on compact groups has revealed that galaxies show a canyon between quiescent and active galaxies in mid-infrared (\mir) colorspace which is unique to the compact group environment \citep{johnson07, gallagher08,walker10,walker12}. This dearth is also seen in the specific star formation rates of galaxies in compact groups \citep{tzanavaris10,lenkic15}. Though likely a transition region, the \mir canyon does not correspond to the optical ``green valley'' between star-forming and quiescent galaxies; in fact these \mir transition galaxies fall on the optical red sequence $\left(0.73 < \left[g-r\right] < 0.77\right)$ \citep{walker13}. The role of gas processing is key to this transformation. \citet{cluver13} found extreme $\mathrm{H_2}$ excitation in compact group galaxies in or near the canyon, further evidence that changing gas properties rapidly extinguish star formation. It appears as though the \mir canyon may result from interactions enhancing star formation, followed by rapid truncation due to a lack of fuel. Unlike simple two-body mergers, the fuel depletion might be enhanced by a variety of processes, including stripping or heating. As the environment in compact groups is similar to those of the earlier universe \citep{baron87}, this has implications for galaxy evolution at higher redshift. To determine the cause of the rapid truncation of star formation in this high density environment we must ``follow the gas'' in compact groups. Determining the dominant mechanisms that both trigger and abruptly quench star formation in compact groups will translate to a deeper understanding of galaxy evolution in the earlier universe.

Understanding galaxy evolution in the dense compact group environment requires not only characterizing each galaxy, but also the overall group properties and dynamics. Gas in all its phases in these systems is key--it provides fuel for star formation, responds to shocking and dissipation, and influences the resulting galaxies. \citet{desjardins13} find that diffuse X-ray emission tracing ionized gas in the majority of compact groups is associated with individual galaxies, rather than the overall group. However, \citet{desjardins14} note that the hot gas in higher-mass systems appears to be in a common envelope. Like clusters, compact groups that have lower \ion{H}{1} content have a higher rate of detection in X-ray \citep{giovanelli85,verdes01}. Single-dish studies of the atomic gas reveal the complexity of compact groups: as a class, compact groups are generally \ion{H}{1}-deficient \citep[as compared with their predicted \ion{H}{1} mass from optical luminosities based on a sample of isolated galaxies;][]{verdes01}; and additionally span a range of deficiencies. In this paper, we expand previous studies of \ion{H}{1} in 73 HCGs \citep[e.g.,][]{verdes01,borthakur10} to Redshift Survey Compact Groups (RSCGs) to form a more comprehensive study of the neutral gas in compact groups, in order to compare with member galaxy properties at other wavelengths. We have obtained high-quality spectra of 30 RSCGs with the 100m Robert C. Byrd Green Bank Telescope (GBT). We present group \ion{H}{1} masses, and compare these with stellar masses, galaxy morphology, and optical and \mir colors.

\section{DATA}
\subsection{Sample}
RSCGs were selected by \citet{barton96} using a friends-of-friends algorithm to identify groups similar to Hickson Compact Groups (HCGs), which were selected by eye from Palomar Sky Survey plates \citep{hickson82} and later confirmed spectroscopically \citep{hickson92}. For this study, we have selected RSCGs at a redshift less than $z = 0.035$ \citep[to match the sample from][]{walker12}. Only a subset of these were actually observed, based on RA availability; these 30 groups are listed in Table \ref{tab:groups}. In our analysis, we also utilize a sample of 73 HCGs with \ion{H}{1} masses from the literature \citep{verdes01,borthakur10}. As \citet{barton96} searched for HCG-like groups, we can interpret the \ion{H}{1} data in the same way as in previous studies. This yields a total sample size of 62 groups with both \ion{H}{1} and stellar data.

\subsection{\ion{H}{1} Data}
New 21 cm \ion{H}{1} observations were obtained on the GBT for 30 RSCGs in June and July 2009 and February, March, April, and October 2010 through the GBT09B-011, GBT10A-050, and GBT10C-022 programs. The observations utilized the L-band receiver with 9-level sampling and a 50MHz ($>$10000 km s$^{-1}$) bandwidth, yielding 16384 channels at 3.052 kHz (0.631 km s$^{-1}$) resolution. The observations were obtained before individual group members had been identified, thus the pointings were chosen using published group RA and Dec values \citep{barton96}; based on the individual galaxy positions some of them are not ideal.

The observations were taken using position switching with two different off positions to be able to reject a bad off location; our sequence was OFF1-ON-ON-OFF2, with five minutes spent at each position. At the beginning of each run, we performed an AutoPeakFocus to minimize pointing errors. The flux calibrators 3C48 (16.2 Jy), 3C249.1 (2.26 Jy), and 3C286 (14.6 Jy) gave an average antenna gain of 1.70 K/Jy \citep{ott94}.

Reduction was performed using GBTIDL \citep{marganian06}. All scans were inspected and scans with radio-frequency interference (RFI) or obviously discrepant baselines were flagged. When RFI was confined to specific channels, those channels were removed in all scans for that group. The calibrated spectra were produced for each polarization/scan. These spectra were combined, weighted by exposure time and system temperature, and then scaled by the K/Jy conversion factor derived from observations of the calibration source. The combined spectra for each group were then boxcar smoothed by 20 channels to yield a 12.6 km/s resolution. Finally, a polynominal fit to the baseline was subtracted to produced a final spectrum.

In most cases, a polynominal fit of order 15 or less was adequate to fit the baseline. However, a subset of 10 groups (RSCG 5, 7, 10, 11, 15, 17, 21, 61, 67, 86) have a 0.02K ripple with a period of 5.5 MHz in their baseline. This ripple is likely due to strong continuum emission. Seven of the affected groups have strong continuum emission in the NVSS \citep{condon98}: RSCG 5, 10, 11, 17, 61, 67, and 86. Of the other three affected groups, two are near 3C48 (RSCG 7 and 15) and RSCG 21, our closest source to the Galactic Plane (b=-13), is near a strong continuum source in the NVSS survey. We attempted to correct these baselines using the double-position switching technique \citep{ghosh02}, but were not successful. While the double-position switching removed some of the structure, it was not able to remove the 5.5MHz ripple, most likely because of time-dependent issues.  Given that we are attempting a line detection, we did not attempt to fit the ripple with a sinsusoid because of the likelihood of spurious detections if the ripple was not perfectly sinusoidal. The resulting noise limits for these groups are due to the systematic effects of the ripple rather than the thermal noise.

\section{RESULTS}
The H I fluxes for each group were measured by summing emission via the GBTIDL procedure gmoment. We detect H I in 15 groups, shown in Figure 1, a detection rate of 50\%. The detection rate may be artificially low because the higher noises introduced by the 5.5MHz baseline ripple. As in \citet{verdes01} and \citet{borthakur10}, fluxes were converted to \ion{H}{1} masses using
$M_\hone [M_\sun] = 2.36\times10^5 \left(\frac{D}{\mathrm{Mpc}}\right)^2 \int \left(\frac{S_\nu}{\mathrm{Jy}}\right)\left(\frac{dv}{\mathrm{km\, s^{-1}}}\right)$. 
Masses and 5$\sigma$ upper limits (using a line width of 400 km/s; the average width from our detections) are listed in Table \ref{tab:masses}.

For some of the groups, we can use the optical radial velocities from NED to identify from which galaxy the emission is coming; e.g., in RSCG 38, the weaker emission line is clearly coming from NGC 3413. Some groups have \ion{H}{1} profiles that do not allow definitive association of emission with a particular galaxy, but it is possible to identify the likely source of the emission; e.g., in  RSCG 34, the majority of the \ion{H}{1} emission is likely coming from NGC 2964, based on the galaxy velocities and \mir colors. However, many of the groups show emission that cannot be identified as coming from a single galaxy, as the velocities of multiple (or even all) member galaxies fall within the \ion{H}{1} profile. In these cases, either the profiles of emission from multiple galaxies overlap, or the \ion{H}{1} is coming from an envelope common to multiple galaxies. Notes on individual groups are given in the appendix.

\subsection{Group \ion{H}{1} Content: Comparison with Stellar Masses}
It is important to place the \ion{H}{1} mass in context with the size of the group. Previous works have compared the group \ion{H}{1} mass to the group dynamical mass to determine \ion{H}{1} content. However, as many of these groups have only 3 known members, it is difficult to determine an accurate dynamical mass to use to calculate content \citep{mcconnachie08}. Another method commonly used is to determine the predicted \ion{H}{1} mass based on a galaxy's luminosity, and sum the masses of the members to determine a predicted \ion{H}{1} mass for the group. However, the optical bands can be highly biased by young, bright stars from recent star formation, while stellar masses are are more robust against recent star formation. Thus, we compute group \ion{H}{1} content using {\it group} stellar masses, which is simply the sum of the stellar masses of the giant members. We define the \ion{H}{1} content as the neutral-gas-to-stellar-mass ratio, $\log{\left(\frac{M_{\hone}}{M_*}\right)}$.

Stellar masses were computed one of two ways, as in \citet{desjardins14}. When possible, they were determined through SED-fitting of Two Micron All Sky Survey (2MASS) $JHK_s$ and \spit IRAC 3.6-8.0 $\mu m$ data using the GRASIL model galaxy spectral library \citep{silva98, silva99, granato00, bressan02, silva03, panuzzo03, vega05, silva09}. When these data were not available, stellar masses were computed using $K_s$-band luminosities and a mass-to-light ratio of 0.95 \citep[\citet{bell03}; for more details see ][]{desjardins14}.

A comparison of the stellar versus \ion{H}{1} mass for both HCGs and RSCGs is shown in Figure \ref{fig:histellar}. The \ion{H}{1} mass is always less than the stellar mass of the group, but the difference is not a constant. The \ion{H}{1} content for the RSCGs and the HCGs is similar as shown in Figures \ref{fig:histellar} and \ref{fig:hicontent}; the median \ion{H}{1} content for the RSCGs is \remind{$\log{\left(\frac{M_{\hone}}{M_*}\right)} = -2.10$, ranging from -2.44 at the 25th percentile to -1.66 at the 75th percentile} and the median \ion{H}{1} content for the HCGs is \remind{$\log{\left(\frac{M_{\hone}}{M_*}\right)} = -1.53$, ranging from -1.97 at the 25th percentile to -1.12 at the 75th percentile}; the two samples are indistinguishable with a KS test, where, with a probability of 27\%, we cannot exclude the hypothesis that they are drawn from the same parent distribution. This provides circumstantial evidence that the different selection criteria for these samples is not affecting the global gas processing in a significant way. Figure \ref{fig:hicontent} also allows us to classify the groups by their \ion{H}{1} content based on the shape of the distribution of groups in the histogram. The dashed lines indicate the boundaries we have defined between \hi-poor/intermediate/\hi-rich groups at $\log{\left(\frac{M_{\hone}}{M_*}\right)} =$ \lobound and \hibound.

\section{DISCUSSION}

\subsection{Morphologies}
Figure \ref{fig:himorph} shows the \ion{H}{1} properties as a function of the spiral fraction of the group. This shows that there is no correspondence between $M_{\hone}$ and spiral fraction. The \ion{H}{1} content, $\log{\left(\frac{M_{\hone}}{M_*}\right)}$, of the group is higher for groups with a larger percentage of spirals and/or irregulars. The large majority of the non-detections are in groups with no spirals. These results make sense, as late-type galaxies tend to be more \ion{H}{1}-rich than early-type (elliptical or lenticular) galaxies.

\subsection{Galaxy \mirc and Optical Colors}
To understand the impact of the group environment, it is important to investigate how {\it galaxy} properties are affected by {\it group} \ion{H}{1} content. Figure \ref{fig:mir} shows the \mir colors of galaxies in compact groups as a function of group \ion{H}{1} content. As in \citet{johnson07}, we see a correlation between group \ion{H}{1} content and galaxy \mir color. Galaxies in \ion{H}{1}-rich groups tend to be actively star-forming, while galaxies in \ion{H}{1}-poor groups tend to be \mir quiescent. This is not surprising, as groups with more \ion{H}{1} have more raw fuel available for star formation. Figure \ref{fig:opt} shows galaxies in the optical color-magnitude diagram (CMD), binned by the \ion{H}{1} content of their group. As with the \mir colors, there is a correlation with optical colors. Galaxies in \ion{H}{1}-poor groups occupy the optical CMD similarly to galaxies in high-density environments (i.e., clusters) with a large red sequence. Galaxies in \ion{H}{1}-rich groups occupy the optical CMD more similarly to galaxies in low-density regions (i.e., the field) with galaxies both in the blue cloud of active star formation and on the red sequence of either quiescence or dust extinction \citep{hogg04}.

\subsection{Rogue Galaxies}
While overall the \ion{H}{1} in most groups was not surprising given the galaxy morphologies, there are three groups with individual galaxies whose \mir and \ion{H}{1} properties are not consistent. The resolved \ion{H}{1} map of RSCG 34 by \citet{tyson01} shows \ion{H}{1} in NGC 2964 and NGC 2968 as well as in the intragroup medium. However, NGC 2968 is quiescent in the \mir; despite the \ion{H}{1} present in the galaxy, it shows no active star formation. KUG 1131+202A in RSCG 42 and NGC 4613 in RSCG 64 are the opposite -- despite the fact that there does not seem to be \ion{H}{1} associated with these galaxies, their \mir colors indicate that they have active star formation occurring. A {\it Swift} study by \citet{lenkic15} show that the UV colors of KUG 1131+202A in RSCG 42 and NGC 4613 in RSCG 64 are consistent with star-forming galaxies and the UV color of NGC 2968 in RSCG 34 is ambiguous.

These three galaxies may be examples of the enhanced evolution experienced by compact group galaxies. The \ion{H}{1} in NGC 2968 has likely been stirred up by interactions such that it cannot currently form stars. The optical spectrum from the Sloan Digital Sky Survey Data Release 12 \citep{alam15} shows hydrogen lines in absorption, indicating the existence of a stellar population less than 1 Gyr old. Conversely, KUG 1131+202A and NGC 4613 may have just had their gas stripped as a result of being in a compact group and their star formation is about to end. None of these galaxies are currently \mir transition galaxies, but if their star formation changes, this could move them into the \mir transition region.

\section{Conclusions}
The \ion{H}{1} properties of our detected RSCGs are similar to those of the HCGs. For all groups, total stellar mass is larger than total \ion{H}{1} mass with a median \ion{H}{1} content of \remind{$\log{\left(\frac{M_{\hone}}{M_*}\right)} = -1.82$, ranging from -2.26 at the 25th percentile to -1.16 at the 75th percentile}. Groups with a smaller fraction of late-type galaxies, on average, have lower \ion{H}{1} content. Galaxies in \ion{H}{1}-rich compact groups occupy the optical CMD similarly to those in low-density environments, where galaxies tend to have a larger reservoir of gas and have more galaxies with \mir colors indicative of activity. Meanwhile, galaxies in \ion{H}{1}-poor compact groups predominantly have quiescent \mir colors and fall on the red sequence -- similar to high-density environments -- where galaxies tend to be depleted of cold gas.

\begin{figure}
  \plottwo{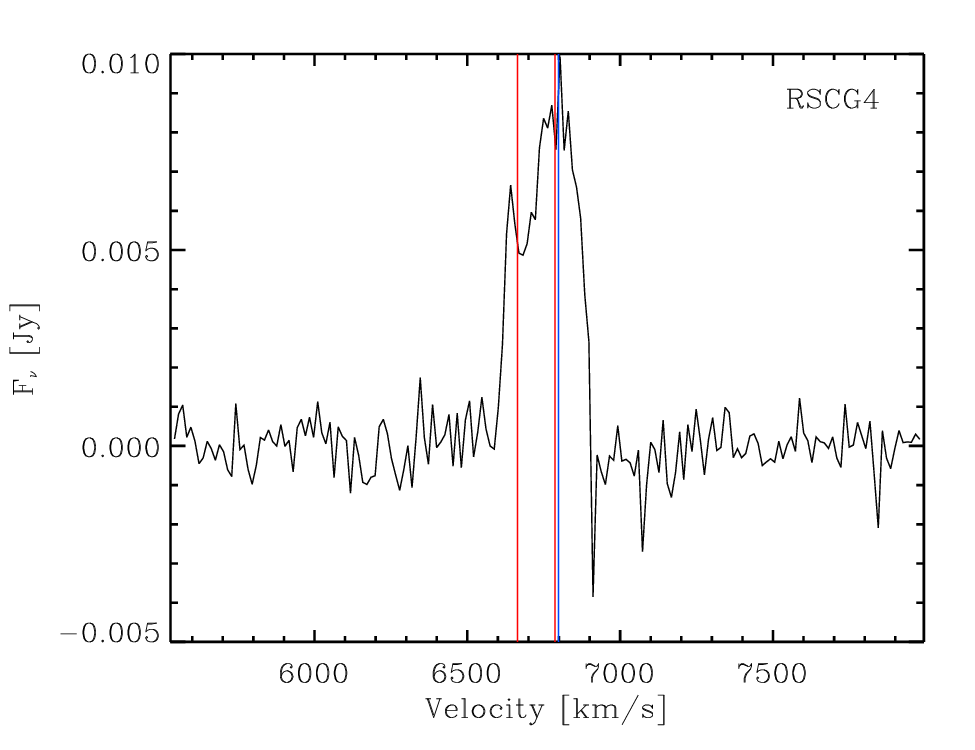}{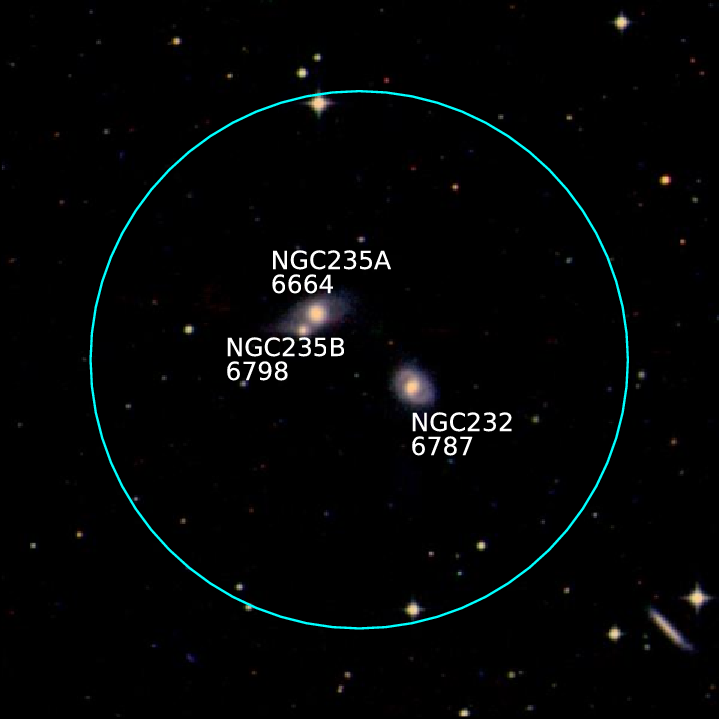}
  \caption{RSCG 4. This \hi-intermediate group has an \ion{H}{1} content of -2.19. {\it Left}: The \ion{H}{1} spectrum. Vertical lines indicate the velocities of member galaxies; line color indicates the galaxy's \mir color--red lines are \mir active, blue lines are \mir quiescent, and black lines are for galaxies without \mir data. {\it Right}: Optical image from SDSS (excepting RSCG 4 and RSCG 16, which are outside the SDSS footprint; these images are from the Digitized Sky Survey) with galaxy velocities labelled in white and the GBT FWHM shown in cyan. It is possible to detect \ion{H}{1} outside this circle; this is simply the area over which the sensitivity of the GBT is greatest. \label{fig:spectra}}
\end{figure}

\begin{figure}
  \plotone{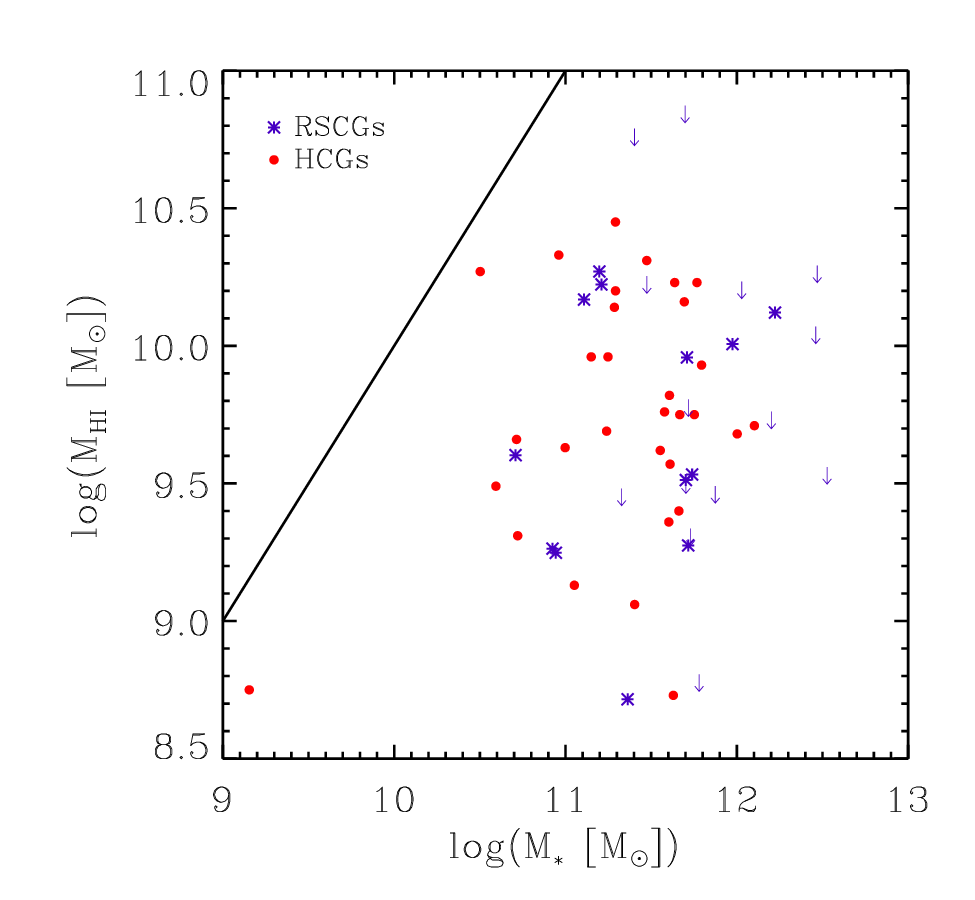}
  \caption[Comparison of $M_{\hone}$ and $M_*$ for compact groups]{Comparison of $M_{\hone}$ and $M_*$; red circles are HCGs, purple asterisks are RSCGs, and purple arrows are RSCG upper limits. Though these are only groups for which both \ion{H}{1} and stellar masses were available, we see that HCGs and RSCGs have similar distributions. The solid line indicates a one-to-one correspondence. Not surprisingly, the compact groups all fall below this line, indicating that they contain more stellar mass than \ion{H}{1}, but with a Spearman's rank correlation coefficient of \remind{0.09} for the detections (\remind{0.10} when including upper limits), there is no significant correlation between $M_{\hone}$ and $M_*$.\label{fig:histellar}}
\end{figure}

\begin{figure}
  \plottwo{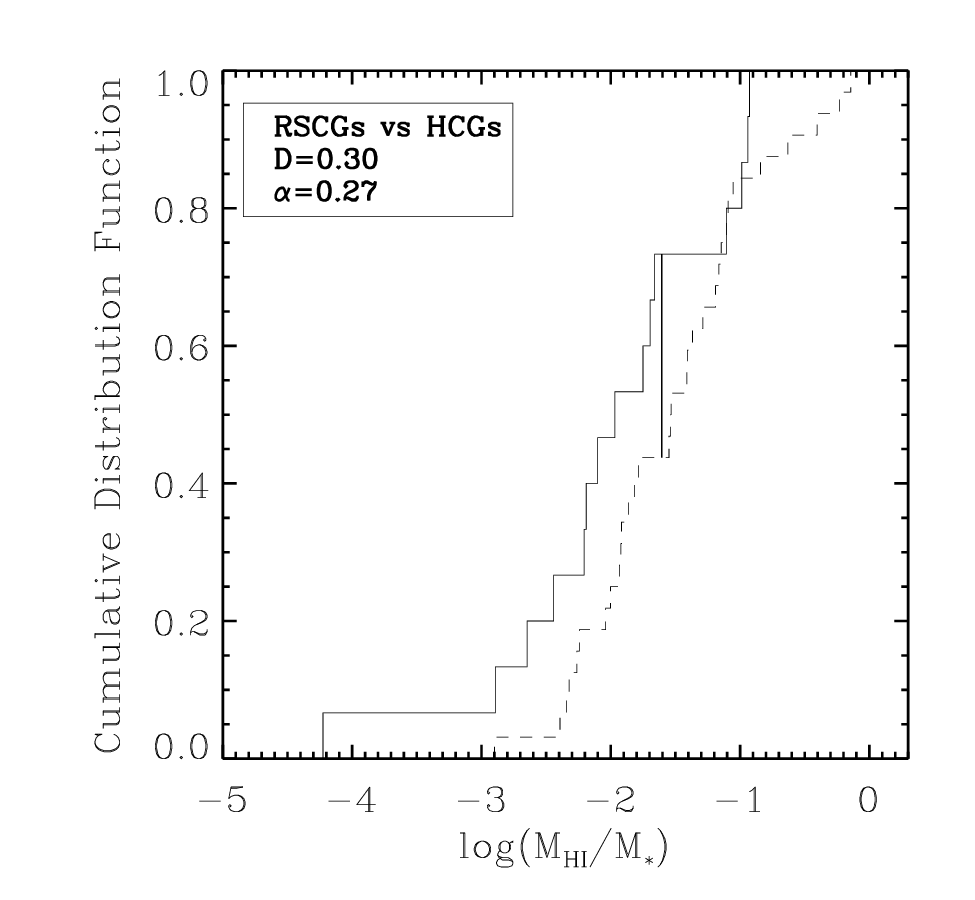}{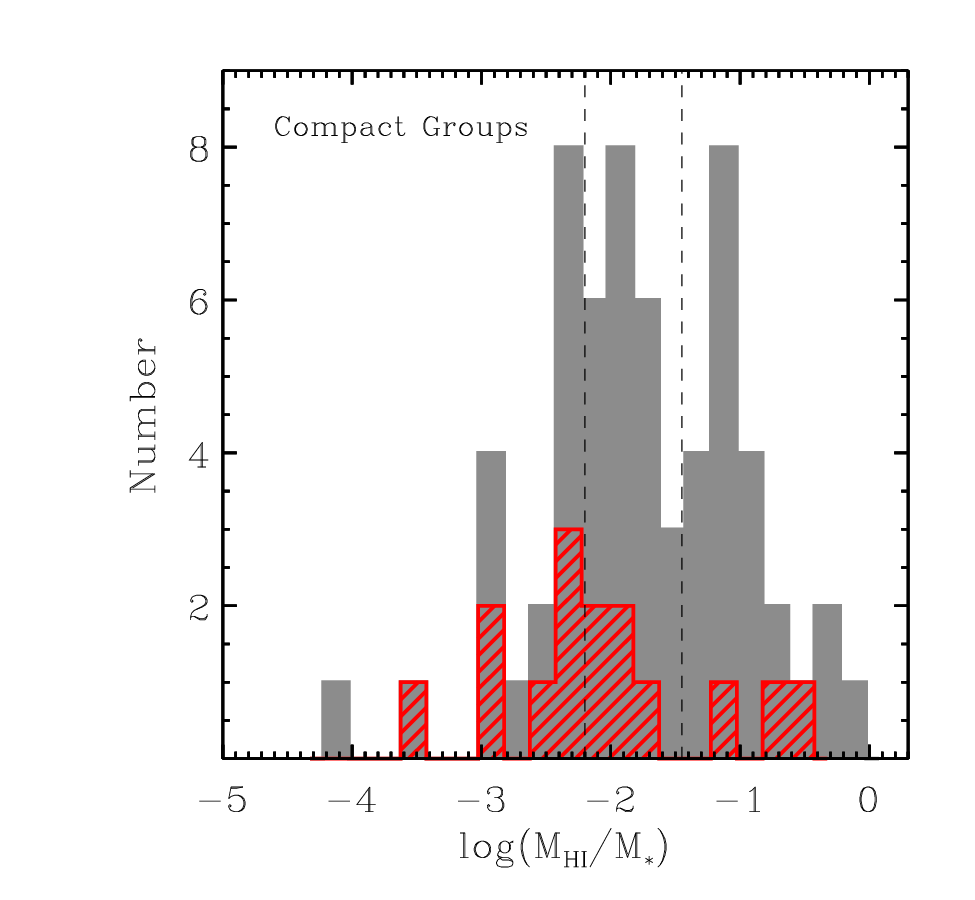}
  \caption[Effect of sample selection on \ion{H}{1}\hspace{0.1em} properties of compact groups]{{\it Left}: Kolmogorov-Smirnov (KS) test comparing $\log{\left(\frac{M_{\hone}}{M_*}\right)}$ for the two samples; the solid line indicates the RSCGs while the dotted line represents the HCGs. As the samples are indistinguishable by the KS test, we combine them for subsequent analysis. {\it Right}: Histogram of $\log{\left(\frac{M_{\hone}}{M_*}\right)}$ of HCGs and RSCGs (upper limits are shown as red subset); the dashed lines show our divisions between \hi-poor/intermediate/\hi-rich at \lobound and \hibound derived from this histogram.\label{fig:hicontent}}
\end{figure}

\begin{figure}
  \plottwo{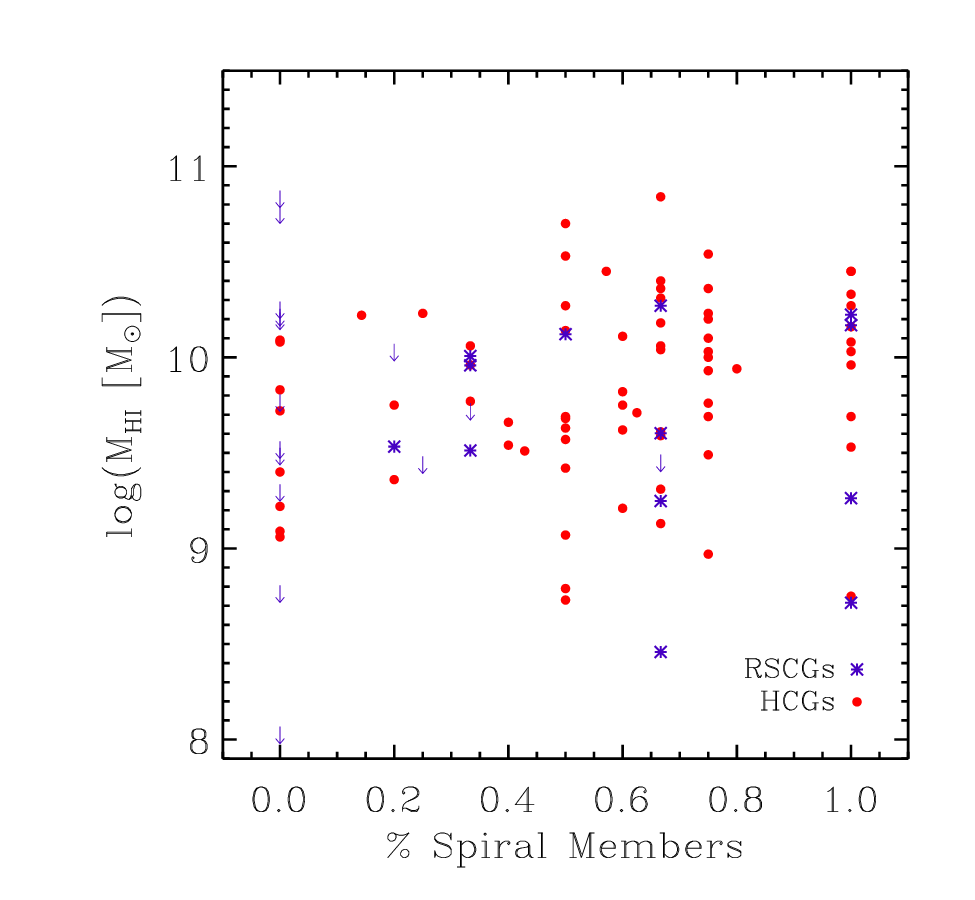}{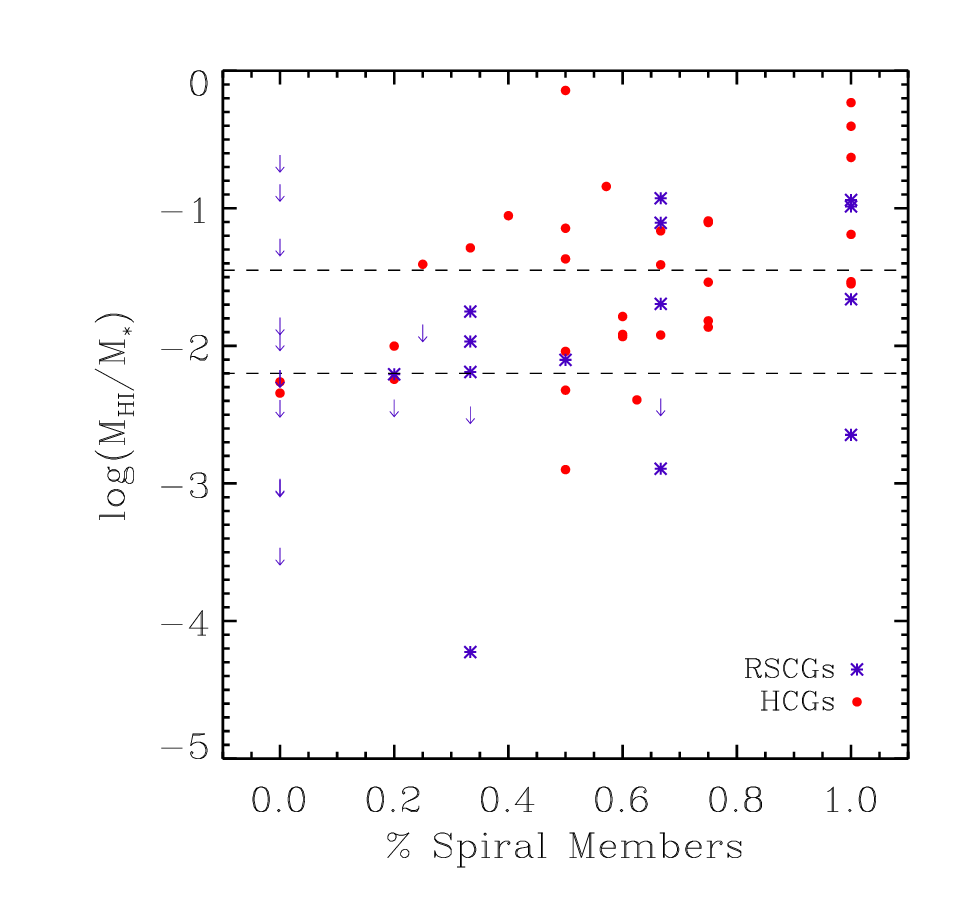}
  \caption[Effect of spiral fraction on \ion{H}{1}\hspace{0.1em} properties of compact groups]{{\it Left}: $\log{\left(M_{\hone}\right)}$ and {\it Right}: $\log{\left(\frac{M_{\hone}}{M_*}\right)}$ as a function of spiral fraction. The dashed lines indicate the divisions between \hi-poor/intermediate/\hi-rich at \lobound and \hibound. In both plots, the red circles are HCGs, purple asterisks are RSCGs. There is no significant correspondence between $M_{\hone}$ and spiral fraction. However, $\log{\left(\frac{M_{\hone}}{M_*}\right)}$ does increase with increasing spiral fraction, albeit with large scatter. Also, note that the distribution of \ion{H}{1} upper limits lies toward lower spiral fractions.\label{fig:himorph}}
\end{figure}

\begin{figure}
  \plottwo{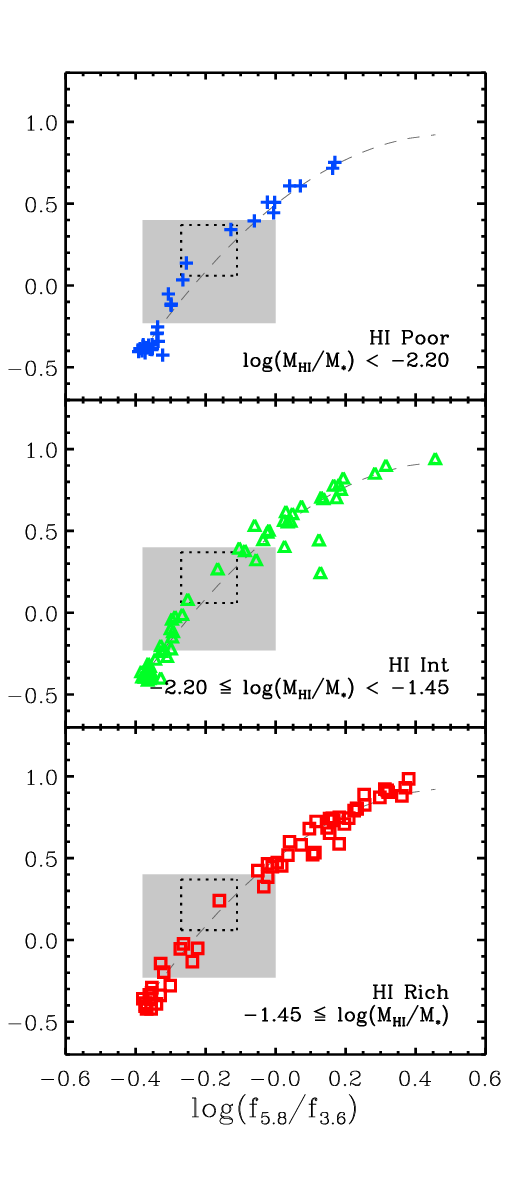}{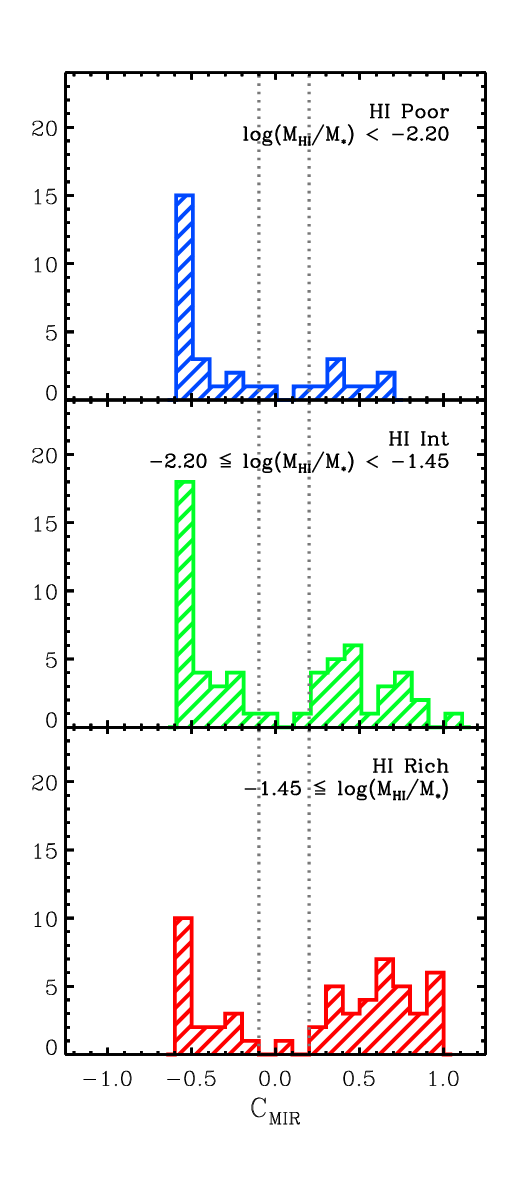}
  \caption[Group \ion{H}{1}\hspace{0.1em} content and galaxy \mir colors]{{\it Left}: The distribution of galaxies in IRAC colorspace, broken up by group \ion{H}{1} content, with the \mir canyon \citep{walker12} and \mir gap \citep{walker10} indicated by the dashed and grey boxes, respectively. {\it Right}: Histograms of \mir color along the dashed line, with active galaxies toward the right. There is a clear correlation between {\it group} \ion{H}{1} content and {\it galaxy} \mir color--a larger fraction of galaxies in \ion{H}{1}-rich groups have colors indicative of activity.\label{fig:mir}}
\end{figure}

\begin{figure}
  \plottwo{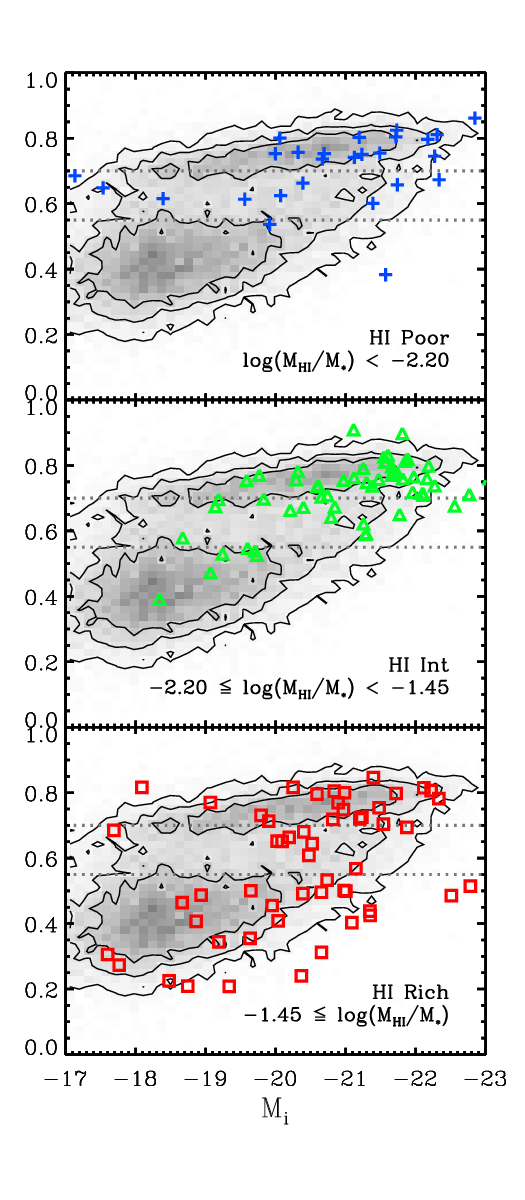}{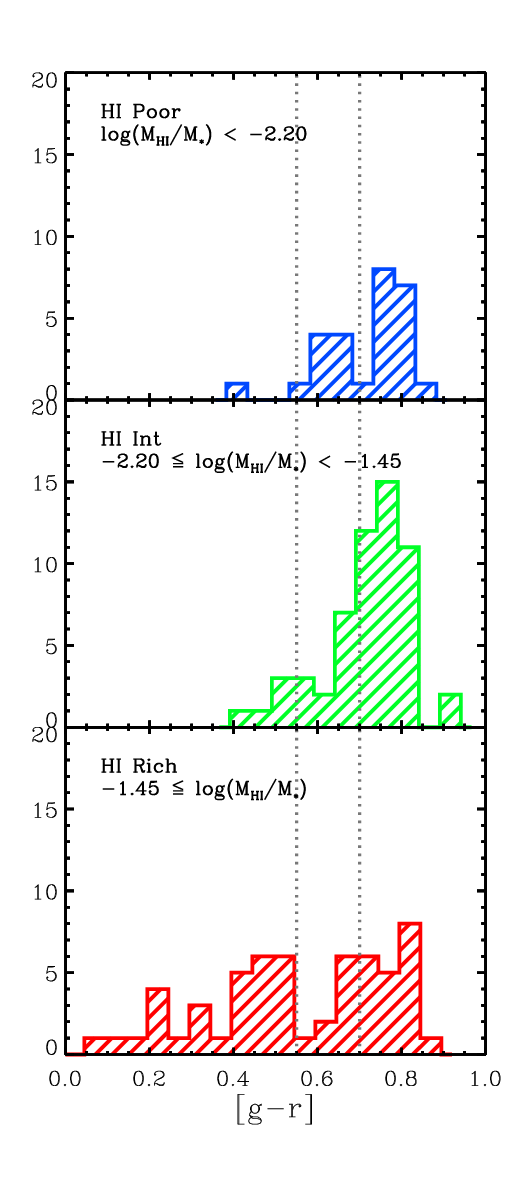}
  \caption[Group \ion{H}{1}\hspace{0.1em} content and galaxy optical colors]{{\it Left}: The distribution of galaxies in the optical CMD, broken up by group \ion{H}{1} content overlaid on the NYU-VAGC \citep[contours and grayscale]{blanton05a}. {\it Right}: Histograms of optical color, where active galaxies can span the full range of colors due to extinction. Like the \mir colors, there is a correlation between {\it group} \ion{H}{1} content and {\it galaxy} optical color. This is reminiscent of how the distribution of galaxies in a CMD changes with galaxy environment. \label{fig:opt}}
\end{figure}

As Figure \ref{fig:spectra} shows, higher resolution interferometer data are required to get a handle on the kinematics in these groups or determine the distribution of \ion{H}{1} -- we do not know whether the \ion{H}{1} is in the galaxies or the intragroup medium. Interferometric observations would also help us measure the HI emission in the seven groups with strong radio continuum emission. To truly understand what is happening to the \ion{H}{1}, we need to map its distribution and compare this with the star formation \citep[from][]{tzanavaris10,lenkic15} and stellar populations of the member galaxies.

We find rogue galaxies in three compact groups whose \ion{H}{1} and star formation are inconsistent. This is likely due to the strong interactions in the compact group environment influencing the \ion{H}{1} and star formation in member galaxies.

\appendix
\section{Notes on Detected Groups}
\groupsec{RSCG 4}
This is a complicated profile with multiple peaks; all three group galaxies are within the GBT beam, and \ion{H}{1} is detected at velocities consistent with the optical radial velocities of each. Without mapping, it is unclear whether the \ion{H}{1} is isolated within individual galaxies or in a common envelope.

The \mir colors reveal that NGC 232 is quite active, NGC 235A is mildly active, and NGC 235B is quiescent. The \mir colors imply that the majority of the \ion{H}{1} emission is likely coming from NGC 232, with some but not as much coming from NGC 235A.

\groupsec{RSCG 6}
RSCG 6 has an asymmetric line profile. Like RSCG 4, all the galaxies are within the beam and the velocities all fall within the line profile, thus the GBT spectrum does not provide enough information to assign emission to individual galaxies.

UGC 813 and UGC 816 are very active in the \mir and CGCG 551-011 is slightly more active than the \mir canyon. From the \ion{H}{1} map in \citet{condon02}, there is a significant amount of \ion{H}{1} in a common envelope surrounding UGC 813 and UGC 816, as well as a small amount in CGCG 551-011.

\groupsec{RSCG 14}
The line profile of RSCG 14 is complex with a main peak at the low velocity side of the spectrum and plateaus of decreasing intensity on the higher-velocity side. Combined with the fact that all three galaxies are in the beam and their velocities are within the line profile, this means that we cannot distinguish the origin of the \ion{H}{1} emission from the GBT spectrum.

Imaging from \citet{vanmoorsel88} shows that edge-on spiral NGC 678 appears to be in a tidal interaction with NGC 680. There appears to be no emission from IC 1730.

\groupsec{RSCG 16}
The line profile is interesting, with three peaks. All three galaxies are within the GBT beam, but the velocity of CGCG 522-061 is quite far from the emission profile. Based on the velocity profile, the \ion{H}{1} emission is most likely associated with UGC 1385 and KUG 152+366.

\groupsec{RSCG 30}
This is another complicated profile that ramps up to a peak on the low velocity side and falls off abruptly after the peak. All three galaxies lie within the beam and their velocities fall within the profile. The velocities are close enough that it is not possible to determine the distribution of the \ion{H}{1} without interferometry.

\groupsec{RSCG 31}
This profile is quite interesting, as the velocities of all three galaxies fall within the profile (and the galaxies are all within the beam), but there is a secondary peak in the profile at a higher velocity than the three member galaxies.

From the \ion{H}{1} mapping in \citet{nordgren97b}, NGC 2799 and NGC 2798 live in a common \ion{H}{1} envelope, and the high-velocity peak seen in the GBT spectrum is associated with the eastern emission of NGC 2799. UGC 4904 has its own \ion{H}{1} cloud in what looks like a slightly morphologically disturbed disk. NGC 2799 and UGC 4904 are both \mir active; NGC 2798 would likely be classified as active, but it is saturated in \mir, so the colors have not been measured.

\groupsec{RSCG 32}
All 3 galaxies are in the GBT beam, but the \ion{H}{1} emission is clearly centered on the velocity of NGC 2830; the velocities of NGC 2831 and NGC 2832 are quite separated from the line profile.

The only \mir-active galaxy is NGC 2830, an edge-on spiral, and the \ion{H}{1} emission is centered at its velocity. Thus the \ion{H}{1} emission is almost certainly coming from NGC 2830.

\groupsec{RSCG 34}
The unusual RSCG 34 has an asymmetric profile centered on the velocity of NGC 2964, which actually lies outside the half-power point of the GBT beam. NGC 2970 also lies outside the higher sensitivity portion of the beam, and the second peak is centered on its velocity. However, the velocity of NGC 2968 is close enough that it could be associated with the emission from the smaller peak.

As seen in \ion{H}{1} imaging from \citet{tyson01}, NGC 2964 and NGC 2968 contain \ion{H}{1} with a bridge connecting the two and a tail stretching towards NGC 2970, but only NGC 2964 is active in the \mir.

\groupsec{RSCG 38}
RSCG 36, like RSCG 34, is also larger than the beam. NGC 3424 is the only galaxy that falls within the GBT beam. The main profile is at the same velocity as both NGC 3424 and NGC 3430, and is asymmetrical. The smaller profile, at the velocity of NGC 3413, indicates a lower-limit to the \ion{H}{1} in NGC 3413 because we are likely missing flux from the galaxy.

NGC 3424 and NGC 3430 are very active in \mir, NGC 3413 is slightly more active than the canyon. NGC 3424 and NGC 3430 each have their own \ion{H}{1} emission that spatially overlap into an envelope \citep{nordgren97a}. The \ion{H}{1} emission coming from NGC 3413 seen in the single-dish spectrum is not apparent in the map.

\groupsec{RSCG 42}
The line profile of RSCG 42 has several parts, and the velocities of the galaxies are all within the profile. All the galaxies are within the beam, thus it is not possible to determine from which galaxy the emission originates.

The \mir colors show UGC 6583 and KUG 1134+202A to be quite active while ARK 303 is very quiescent. The \ion{H}{1} map in \citep{scott12} show that there is a common \ion{H}{1} envelope surrounding all three which extends westward of group, with the peak of the emission coming from UGC 6583. Despite being within the \ion{H}{1} cloud in projection, there is no peak associated with KUG 1134+202A, which is surprising as it is active in the \mir.

\groupsec{RSCG 44}
All the galaxies in RSCG 44 are within the GBT beam. However, the  quite asymmetrical emission is centered at the velocity of UGC 6697; the velocities of the other galaxies do not overlap with the line profile.

The \mir data show that UGC 6697 is fairly active and the other four galaxies are quiescent. Interferometric observations in \citet{scott10} indicate that the \ion{H}{1} is associated with UGC 6697, which is located near a peak in X-ray gas in a subcluster in Abell 1367. This galaxy is in the early stages of being ram-pressure stripped \citep{sun05}.

\groupsec{RSCG 53}
The highest sensitivity portion of the GBT beam did not overlap with any of the galaxies in RSCG 53. However, the emission it did detect is highly asymmetrical and overlaps with the velocities of all three galaxies in the group.

VLA observations of NGC 4302 and NGC 4298 detect \ion{H}{1} in these two galaxies \citep{chung09}; it is unclear whether UGC 7436 was in the VLA beam.

\groupsec{RSCG 54}
Though the GBT beam only includes M85, the velocity of the emission is consistent with NGC 4394. The velocities of M85 and IC 3292 are both lower than the velocity of the emission.

An \ion{H}{1} map from \citet{hibbard03} shows that the \ion{H}{1} is contained within NGC 4394.

\groupsec{RSCG 64}
The fairly peaked line profile is centered at the velocity of NGC 4615, though it is not inconsistent with the velocities of NGC 4613 and NGC 4614. All three of these galaxies fall within the GBT beam.

NGC 4613 and NGC 4615 are fairly active in the \mir, while NGC 4614 is fairly quiescent. This implies that the majority of the \ion{H}{1} is probably in NGC 4615, but NGC 4613 needs some fuel in order to form stars. The disturbed optical morphology of NGC 4615 suggests that it has been affected by the compact group environment and has likely had a recent interaction. An \ion{H}{1} map would reveal how the gas has reacted to this and with which galaxy NGC 4615 interacted.

\groupsec{RSCG 78}
The shape of this line profile is quite interesting. The majority of the emission is contained in what looks to be an asymmetric double-horned profile centered at the velocity of NGC 6962. However, at the lower velocities of the other three galaxies in the group, we do see emission. The intriguing origin of the \ion{H}{1} in RSCG 78 cannot be determined without a map.

\acknowledgments
K.E.J. gratefully acknowledges support for this paper provided by NSF through CAREER award 0548103 and the David and Lucile Packard Foundation through a Packard Fellowship. GCP was supported by a FONDECYT Postdoctoral Fellowship (No.\ 3150361). Support for this work was provided by the NSF through award GSSP 10A-003 from the NRAO. The National Radio Astronomy Observatory is a facility of the National Science Foundation operated under cooperative agreement by Associated Universities, Inc. We thank the anonymous referee for their constructive comments. This research has made use of the NASA/IPAC Extragalactic Database (NED) which is operated by the Jet Propulsion Laboratory, California Institute of Technology, under contract with the National Aeronautics and Space Administration. 

{\it Facilities:} \facility{100m Robert C. Byrd Green Bank Telescope (GBT)}.

\input{Groups.tex}
\input{MassesUplim.tex}

\plotmorespec{6}{-0.94}{\hi-rich}
\plotmorespec{14}{-1.97}{\hi-intermediate}
\plotmorespec{16}{-2.44}{\hi-poor}
\plotmorespec{30}{-2.89}{\hi-poor}
\plotmorespec{31}{-1.66}{\hi-intermediate}
\plotmorespec{32}{-1.75}{\hi-intermediate}
\plotmorespec{34}{-1.70}{\hi-intermediate}
\plotmorespec{38}{-1.11}{\hi-rich}
\plotmorespec{42}{-0.93}{\hi-rich}
\plotmorespec{44}{-2.21}{\hi-poor}
\plotmorespec{53}{-2.65}{\hi-poor}
\plotmorespec{54}{-4.23}{\hi-poor}
\plotmorespec{64}{-0.99}{\hi-rich}
\plotmorespec{78}{-2.10}{\hi-intermediate}

\clearpage

\begin{multicols}{2}
{\footnotesize
\bibliography{/home/lisamay/refs/refs}
}
\end{multicols}
\clearpage

\end{document}